\documentclass[manuscript]{aastex6}
\def\etal{et~al.~}

\usepackage{amsmath}
\usepackage{afterpage}
\begin{document}


\title{Prediction of Solar Flares Using Unique Signatures of Magnetic Field Images}

%% Use \author, \affil, plus the \and command to format author and affiliation
%% information.  If done correctly the peer review system will be able to
%% automatically put the author and affiliation information from the manuscript
%% and save the corresponding author the trouble of entering it by hand.
%%
%% The \affil should be used to document primary affiliations and the
%% \altaffil should be used for secondary affiliations, titles, or email.

%% Authors with the same affiliation can be grouped in a single
\author{Abbas Raboonik, Hossein Safari, Nasibe Alipour}
\affil{Department of Physics, University of Zanjan, P.O. Box 45195-313, Zanjan, Iran}

\and

\author{Michael S. Wheatland}
\affil{Sydney Institute for Astronomy, School of Physics, The University of Sydney, NSW 2006, Australia}

%\altaffiltext{1}{safari@znu.ac.ir}
\begin{abstract}

Prediction of solar flares is an important task in solar physics.
The occurrence of solar flares is highly dependent on the
structure and the topology of solar magnetic fields. A
new method for predicting large (M and X class) flares is
presented, which uses machine learning methods applied to the
Zernike moments of magnetograms observed by the
\textit{Helioseismic and Magnetic Imager} (HMI) onboard the
\textit{Solar Dynamics Observatory} (SDO) for a period of six years from 2
June 2010 to 1 August 2016.
Magnetic field images
consisting of the radial component of the magnetic field
are converted to finite sets of Zernike moments and fed to the
Support Vector Machine (SVM) classifier. Zernike moments have the capability
to elicit unique features from any 2-D image, which may
allow more accurate classification. The results indicate
whether an arbitrary active region has
the potential to produce at least one large flare. We show that
the majority of large flares can be predicted within 48 hours
before their occurrence, with only 10 false negatives out of 385 flaring
active region magnetograms, and 21 false positives out of 179 non-flaring active region magnetograms.
Our method may provide a useful tool for prediction of solar flares which
can be employed alongside other forecasting methods.

\end{abstract}




\section{Introduction} \label{sec:intro}

It is accepted that the energy release
 mechanism of solar and
stellar flares is based on magnetic field reconfiguration,
however the exact underlying chain of processes remains ambiguous
\citep{Priest2002}. Accurate forecasting of solar flares is an
extremely important task due to their effect on space weather
\citep{rust1993, Schwenn2006, pulkkinen2007, Wheatland2005, leka2008}. Many
forecasting methods -- e.g. those based on sunspot
classification, time series analysis, avalanche models, machine
learning algorithms, and others -- have been proposed. In recent years the quality and frequency of observations has increased, e.g. due to the availability of data from
\textit{Solar Dynamics Observatory} (SDO) and other satellites. The new data should enable more accurate prediction. However, that requires prediction methods which identify, and take advantage of,  additional information in the data.

McIntosh (1986; 1990) presented a flare forecasting method
named THEO (Theophrastus) which is an expert system
based on sunspot classification. In the extended approach, the
McIntosh classification is primary and some additional
information including magnetic field properties and time series
of former large flares is used \citep{McIntosh1990}.
Wheatland (2004; 2005) investigated a flare prediction method  that exploits solar flare
statistics using Bayesian analysis. In this method, predictions
are made based on the observed time series of flares and the
phenomenological distributions of events in energy and time. The
method was shown to produce forecasts comparable in accuracy to those issued by the National Oceanic and Atmospheric Administration (NOAA), which have been based on THEO (Wheatland 2005).
\cite{belanger2007} applied a four-dimensional variational data assimilation method and
an avalanche model for prediction of large solar flares.
Avalanche models have also been proposed as a basis for solar flare forecasting by
\cite{strugarek2014}, who suggested that such models could lead
to significant improvement in prediction of large solar flares, since solar flares are stochastic in nature. Guerra \etal (2015) presented a method called ``Ensemble Flare Prediction''
in which three flaring probabilities
derived from three different methods used by the Community Coordinated Modeling Center (NASA-GSFC)
and the flare forecasting results provided by the NOAA are linearly combined to five a final flaring probability.

Because of the magnetic origin for large solar energetic events (i.e. flares 
and coronal mass ejections), most large-event prediction methods use 
measured properties of the photospheric vector magnetic field including 
the magnetic flux of ARs (active regions) 
\citep{Kunzel1960,sammis2000,leka2003,Georgoulis2007,Schrijver2007,Falconer2008,Mason2010, Falconer2011, Georgoulis2012, Abramenko2015}. 
\cite{leka2007} used discriminant analysis applied to a set of 
photospheric magnetic quantities, computed from the vector field data 
observed by the University of Hawai‘i Imaging Vector Magnetograph, and 
showed that only a few variables and/or their combinations are related 
to the flare productivity of ARs. \cite{leka2008} debated how the 
performance of different solar flare forecasting methods which 
incorporate different data sources should be compared. They used skill 
scores to compare the ability of those methods that are based on a 
number of parameters computed for photospheric vector magnetic field 
data to forecast the flaring time of large flares. At a flare 
forecasting workshop held in 2009, a variety of prediction methods were 
tested on a common data set. The participating methods were not found to 
perform substantially better than ``climatological'' forecasts, i.e.\ 
predictions based on long-term averages \citep{all-clear}.

Recently, machine learning algorithms have been applied to both 
forecasting of solar flares 
\citep{colak2009,yuan2010,huang2013,yang2013,boucheron2015,shin2016} and 
coronal mass ejections \citep{Bobra2016}. \cite{ahmed2013} developed a 
solar flare prediction method using a feature selection of 21 magnetic 
element properties produced by \textit{Solar Monitor Active Region 
Tracker} (SMART) and a machine learning base classifier. They identified 
that a diminished set of six magnetic features produced a similar 
forecasting results to the whole set of 21 magnetic features. 
\cite{bobra2015} computed 25 quantities from four years of vector 
magnetic field data from 2071 active regions recorded by SDO and examined the relationship 
with flaring. They used the \textit{f}-score feature selection algorithm 
to select the parameters with the highest score. They concluded that 
using four parameters, namely, the total unsigned current helicity, the 
total magnitude of the Lorentz force, the total photospheric magnetic 
free energy density, and the total unsigned vertical current, resulted 
in nearly the same forecasting efficiency as the whole set of 25 
parameters. Using the four parameters listed above and a 
machine learning algorithm, the Support Vector Machine (SVM), they grouped 
ARs into two separate classes. They defined a positive class that 
encompasses all those ARs that will produce at least one large flare 
within a given time interval, and a negative class that contains all 
those ARs that will not produce any flare in the same time interval.

Zernike Moments (ZMs) provide a decomposition of image data which is invariant under scaling, translation, and rotation, and hence in this sense is unique \citep{zernike1934}. These moments have previously been applied, together with machine learning algorithms, to the task of identifying and tracking solar photospheric and coronal bright points and mini-dimmings \citep{alipour2012,alipour2015,javaherian2014}.
In this paper, these methods are adopted as a predictor algorithm
for solar flares. Following the approach of \cite{bobra2015}, magnetograms for ARs are categorized into two
distinct classes, namely, positive and negative, corresponding to
whether the ARs have or have not produced large flares,
respectively. The Zernike moments (ZMs) are calculated for the AR
magnetograms in the two categories. Then, by using a well-trained
machine learning algorithm, we attempt to identify the
corresponding class (positive or negative) for any given AR
magnetogram.
The motivation for implementing the Zernike moments in solar flare forecasting
is to provide a set of unique features for each magnetogram treated as an image.
It is anticipated that this will improve the performance of the classification process in comparison with classifiers trained with just a few global parameters (e.g. total flux, current helicity, etc.) extracted from vector magnetic fields.

The paper is organized as follows: First, the data
processing and the method are discussed in Section 2, and then
the results are given in Section 3. A discussion is presented in
Section 4 followed by an Appendix with additional details of the machine learning methods.

\section{Data Processing and Method} \label{sec:method}

\subsection{Data}

The \textit{Helioseismic and Magnetic Imager} (HMI) instrument
onboard SDO has been returning full-disk solar photospheric vector
magnetic field data since 2010 \citep{schou2012}. In the
present study, we use the Cylindrical Equal Area (CEA) version of the \textit{Spaceweather HMI Active Region Patch} (SHARP) data hmi.sharp\_cea\_720s  (\url{http://jsoc.stanford.edu/ajax/lookdata.html?ds=hmi.sharp\_cea\_720s}), including magnetic field data for 422 National
Oceanic and Atmospheric Administration (NOAA) ARs. The ARs used
were observed in the time period 2 June 2010 to 1 August 2016.
The CEA SHARP vector magnetic data are projections of magnetograms in CCD coordinates onto
heliographic cylindrical equal area coordinates after rotation to disk center.
Here, we use only the radial component of the vector
magnetic field, namely, $B_{r}$. For more information about SHARP vector magnetic field data see \cite{hoeksema2014}.  Using the
 \textit{Geostationary Operational
 Environmental Satellite} (GOES) flare catalogue (\url{ftp://ftp.ngdc.noaa.gov/STP/space-weather/solar-data/solar-features/solar-flares/x-rays/goes/xrs/}) we identify
 113 NOAA ARs, out of the 422 collected ARs, which generate large
 (M and X class) flares during the above mentioned period. Magnetograms dated from 2 June 2010 up to 1 June 2014, and from 1 June 2014 to 1 August 2016 are chosen for the training and test sets, respectively.\\

\subsection{Zernike Moment Representation}

 Zernike Moments are derived from Zernike polynomials (Zernike 1934), which are defined in a unit circle ($x^{2}+y^{2} \leq 1$) and are given in polar coordinates $(r,\theta)$ by:
\begin{equation}\label{eq1}
U_{n,m}(r,{\theta}) = R_{n}^{m}(r) \exp{(im{\theta})},
\end{equation}
where $n$ and $m$ are positive integers, and where $R_n^m(r)$ is given by
\begin{equation}\label{eq2}
R_{n}^{m}(r) = \sum_{s=0}^{\dfrac{_{1}}{^{2}}(n-m)} (-1)^s \, \dfrac{(n-s)!}{s! \, [\dfrac{_{1}}{^{2}}(\small{n}+\small{m})-s]! \, [\dfrac{_{1}}{^{2}}(\small{n}-\small{m})-s]!} \, r^{n-2s}.
\end{equation}
Zernike polynomials have three fundamental properties: they
satisfy orthogonality conditions and form a complete set or
vector space basis; their absolute values are invariant under
rotation; and they force constraints on the $n$ and $m$ indices,
namely $n{\geq}0$ where $n{\geq}{\mid}m{\mid}$, and where $n{\pm}m$ is
an even number.

 Using Zernike polynomials, a 2-D magnetogram
image \textbf{$B_{r}(x,y)$} can be mapped to a complex feature space, but first the image must be transformed from
Cartesian coordinates to polar coordinates. To do this, a 
square magnetogram image is mapped onto a unit disk with the center of
the image mapping to the origin of the polar coordinates. A
thorough explanation about transforming images from Cartesian to
polar coordinates is given by \cite{hosny2010}. The ZMs for the feature space are defined by \citep{hosny2010}:

\begin{equation}\label{eq3}
Z_{n,m} = \dfrac{n + 1}{\pi}\int_{0}^{2\pi}\int_{0}^{1}{U_{n,m}^{*}(r,{\theta}) B_{r}(r,\theta) \, rdr d\theta},
\end{equation}
where the asterisk denotes the complex conjugate.

The magnitudes of ZMs are
invariant under rotation because of the exponential
 angular factor $\exp({im\theta})$ in Equation (\ref{eq1}), but they can also be made invariant under translation and scaling.
  This can be done by transforming an arbitrary image $I(x,y)$
  into a new image $I(x/a + \bar{x} , y/a + \bar{y})$,
  with $\bar{x}$ and $\bar{y}$ being the location of the image
  centroid and $a$ being the scale factor computed from the
  first order normal moments \citep{khotanzad1990,hosny2010}.
With a proper normalization, this produces ZMs which are
invariant to scale. These
properties of ZMs mean that they uniquely characterize any two-variable function. Here we calculate ZMs for magnetogram images of ARs, as a basis for classifying whether the ARs produce large flares (positive class) or do not (negative class).

Figure \ref{fig1} depicts different terms of ZMs for magnetic field data for ARs belonging to the positive class (flare producing) and to the negative class (non-flare producing), respectively.
The figure illustrates how the Zernike moments describe an image. 
The radial part of the Zernike polynomials is bounded to unity ($R_{n}^{m}(r)\le 1$) inside the unit disc. In Equation (\ref{eq3}), the magnetogram
image $B_r$ is weighted by $rR_{n}^{m}(r)$, which is bounded to $r$ inside the unit disc. This means that pixels closer to perimeter of the 
disc have more weight than those closer to the center of the disc. Increasing the polynomial order $n$  leads to an increase in the frequency of oscillations of the polynomial 
along the radial direction. This provides a high capability to describe the details of a magetogram image with a set of ZMs due to the
polynomial oscillation. As we see in Figure \ref{fig1}, the magnitude values of ZMs have different oscillations and shapes for the two magnetograms 
from the
flaring and non-flaring ARs.

Based on the
orthogonality of the Zernike polynomials and using the ZM
coefficients ($Z_{n,m}$) a digital image reconstruction $\widehat{B}_{r}(r,\theta)$ can be made using

\begin{equation} \label{eq4}
\widehat{B}_{r}(r,\theta) = {\sum_{n=0}^{N}}\sum_{\substack{m=0 \\ |n-m|=\rm{even}}}^n{Z_{n,m}U_{n,m}(r,\theta)},
\end{equation}
where ideally, $N$ is infinity.
Using Equation (\ref{eq4}) and a finite number of terms defined by $n\leq{N}$ we can reconstruct the magnetic field image from the ZMs. The optimal value of $N$ is determined empirically, and found to
be 31. This is decided based on the image reconstruction error \citep{javaherian2014}:

\begin{equation}
E^{2}(N) = \dfrac{\sum_{i}\sum_{j}[B_{r}(i,j) - \widehat{B}_{r,N}(i,j)]^{2}}{\sum_{i}\sum_{j}[B_{r}(i,j)]^{2}},
\end{equation}
where $B_{r}(i,j)$ represents
an element of the original magnetogram array and $\widehat{B}_{r,N}(i,j)$ is
an element of the reconstructed magnetogram array,
and the sum is over all possible $i$ and $j$.
The minimum
reconstruction error defines the best value for $N$. In practice this is determined by trial and error.
More information about image reconstruction and associated relative errors is given by
 e.g. \cite{khotanzad1990}, \cite{hosny2010}, and \cite{javaherian2014}. 
 
Figure \ref{fig2} shows an example of a reconstructed image of a positive class magnetogram belonging to the NOAA active region number 11504 on 14 June 2012 at 12:00 UT. It should be noted that there are artefacts and errors in the reconstructed image, so that the two panels in the figure do not fully correspond. One error is due to mapping the original image into polar coordinates and another is due to intrinsic defects in numerical methods \citep{liao1998}. The reconstructed image is not used for the process of classification and is included only to illustrate the image reconstruction process.

 \subsection{Prediction Method}
Here, we propose a flare prediction method using
invariant and unique properties of the Zernike Moments (ZMs), and the Support Vector Machine (SVM) classifier. 
The SVM classifier is a supervised
statistical machine learning method which is based on
Lagrange multiplier optimization \citep{Vapnik1995} and is defined specifically for two-class problems \citep[e.g.][]{gunn1997}. In supervised
learning, the labeled training data set consists of training
examples (pairs of typical vectors in an $l$-dimensional space as the input objects). The SVM classifier attempts to find a separating hyperplane with a maximum margin between
the two classes inside the training set. The maximum margin ensures the least possible error
in classification. The process to find this hyperplane can be simplified to solving an optimization problem (Equation (14) in the Appendix). The SVM code used in this work is the SVM-KM MATLAB toolbox (\url{http://asi.insa-rouen.fr/enseignants/~arakoto/toolbox/SVM-KM.zip}). The regularization parameter $c$ (Equation (14) in the Appendix) is set to $1$ and the kernel function $K(x_{i}, x_{j})$ (Equation (16) in the Appendix) used here
is Gaussian.
After these required procedures, the learning
algorithm can infer (predict) the probable relative class for unseen cases.
 Further details
of the SVM are discussed in the Appendix and also in Tan \etal (2006).\\

\begin{figure}[!ht]
\centering
\centerline{\includegraphics[width=1.2\textwidth]{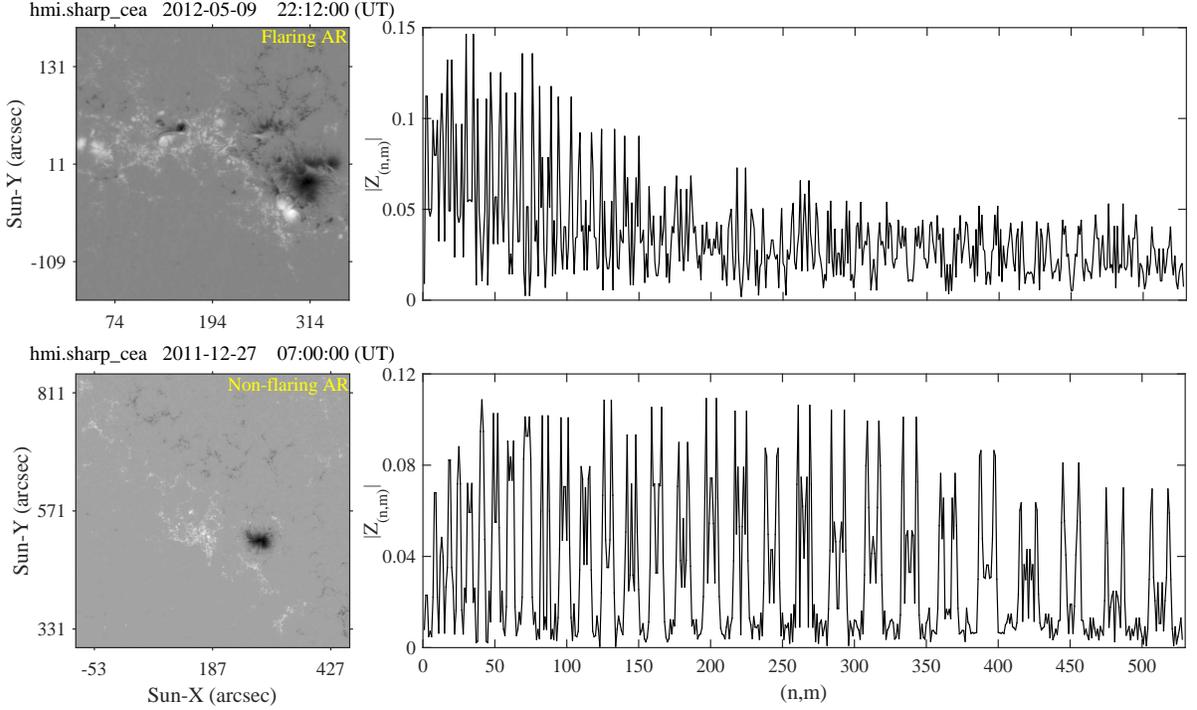}}
\caption{Absolute value of the Zernike
moments versus indices $(n,m)$ for a typical AR in the
positive class (top panel), and a typical AR in the negative
class (bottom panel). Note that each point on the horizontal axis is designated by a pair of integers $(n,m)$ delimited by
the third property of the Zernike polynomials which is $|n \pm m| = \rm{even}$. In the case of $N = 31$ it means that $n$ takes values from 1 through 31. Applying the third confinement rule of the Zernike Polynomials yields 528 pairs of $(n,m)$ in the following way: when $n = 0$ the only possible number for $m$ is 0, if $n = 1$ the acceptable numbers for $m$ are $+1$, $0$, and $-1$, and so on.}\label{fig1}
\end{figure}

\begin{figure}[!ht]
\centering
\centerline{\includegraphics[width=1\textwidth]{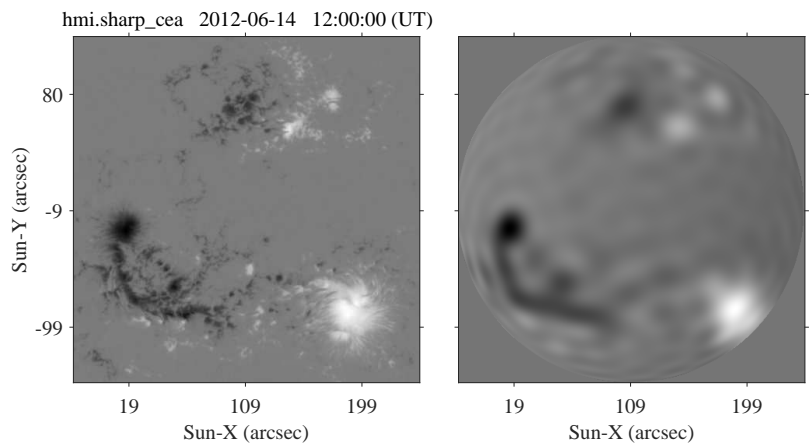}}
\caption{Reconstructed image using the first 528 ($N=31$) Zernike
moment terms for a
magnetogram image of NOAA active region number 11504 on 14 June 2012 at 12:00 UT which is in the positive class. Left: Radial component ($B_{r}$) magnetogram. Right:
Reconstruction of the left-hand side image using the first 528 ZM
terms.}\label{fig2}
\end{figure}

\begin{figure}[!ht]
\centering
\centerline{\includegraphics[width=1\textwidth]{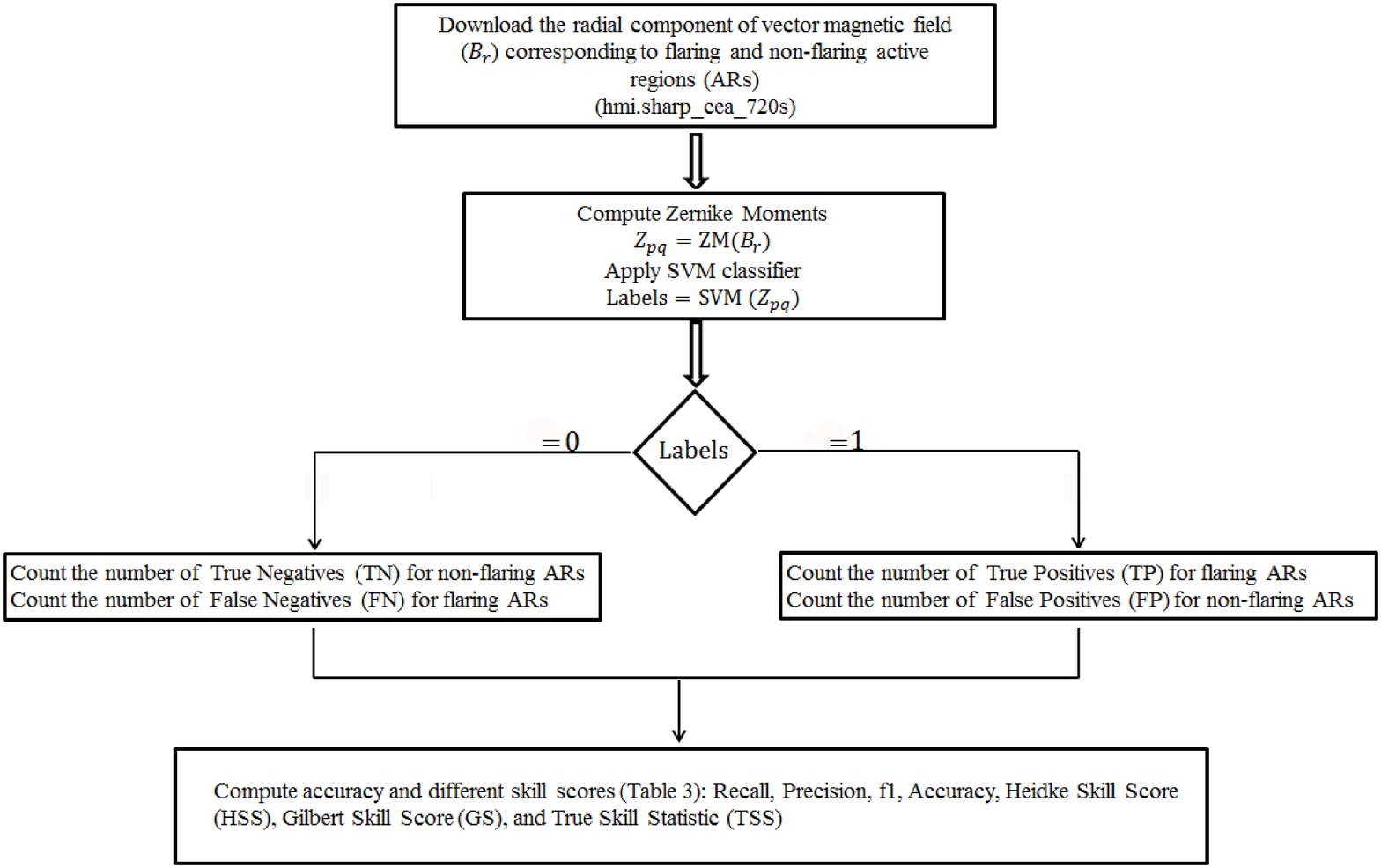}}
\caption{The flowchart of our proposed method.}\label{fig-flowchart}
\end{figure}

As noted above, we divide the magnetograms
for the ARs into two classes, namely a positive and a negative class,
corresponding to all the ARs that produce at least one large flare (M and X class)
within a certain time interval, and those ARs which do not produce
any large flares within the same time interval, respectively. The
ZMs of each magnetogram are  distinctive enough  to be separated
using the SVM classifier, as illustrated in Figure \ref{fig1}. Figure \ref{fig-flowchart} depicts the flowchart of our flare prediction algorithm, for reference.

\section{Results} \label{sec:results}

In this paper, using the unique and invariant properties of ZMs of
the photospheric magnetogram images and the SVM classifier, we
attempt to predict which of the ARs at hand will produce at least
one large flare within 48 hours. We divide the whole
data set into a training and a test set. A supplement to this paper provides electronic tables which contain the ZMs calculated for each magnetogram in the training and test data sets as MATLAB structures, with the exact time and the NOAA AR numbers given for each one. The ARs used in this paper consist in total of 422 different NOAA active regions observed during the time period 2 June 2010 to 1 August 2016. The training set consists of a total of 85 different NOAA ARs belonging to the positive
class observed in the time period 2 June 2010 to 1 June 2014, meaning they produced at least one large flare within 48
hours, and 208 different NOAA ARs belonging to the negative
class observed in the same period, meaning they did not produce a large flare  within the
same time interval. Empirically, in the process of training the positive
class to the SVM we use 6, 2, 2, 2, 2, 2, 2, and 2 magnetogram images (20 in total)
from 1, 5, 18, 20, 22, 24, 25, and 48 hours, respectively before the flaring time of
each of the 85 ARs (Table \ref{tab1}).
Also, the data used to train the negative class to the SVM
consists of almost 7 magnetogram images for each of the 208 ARs
that did not produce any large flare within the past 48 hours.
The
SVM was trained on the ZMs extracted from this data set.
\begin{table}[h]
\centering
\caption{Number of positive class magnetograms corresponding to each time interval before the flaring time used in the training set.}
\label{tab1}
\begin{tabular}{ll}
\hline
\multicolumn{1}{c}{\begin{tabular}[c]{@{}c@{}}Hours before each\\ large flare\end{tabular}} & \multicolumn{1}{c}{\begin{tabular}[c]{@{}c@{}}Number of~~~~~~~~~\\ magnetograms~~~~~~~~~\end{tabular}} \\ \hline
\hspace*{2.1cm}1    & \hspace*{1.73cm}6                      \\ \hline
\hspace*{2.1cm}5    & \hspace*{1.73cm}2                      \\ \hline
\hspace*{2.1cm}18  & \hspace*{1.73cm}2                      \\ \hline
\hspace*{2.1cm}20  & \hspace*{1.73cm}2                      \\ \hline
\hspace*{2.1cm}22  & \hspace*{1.73cm}2                      \\ \hline
\hspace*{2.1cm}24  & \hspace*{1.73cm}2                      \\ \hline
\hspace*{2.1cm}25  & \hspace*{1.73cm}2                      \\ \hline
\hspace*{2.1cm}48  & \hspace*{1.73cm}2                      \\ \hline
\end{tabular}
\end{table}

The rest of the data are taken as the test
set, which consists of 129 ARs. We pretend that we don't know whether these ARs are positive or negative. There are at most 4 magnetogram images in four different times for each the ARs inside the test set. The goal is to identify the corresponding class for every magnetogram image in the test set.

Analysis of the output of the classifier is presented in Table \ref{tab2}, in which TP (True Positive) denotes the number of flaring ARs that are correctly classified as being a member of the positive class (375), FP (False Positive) denotes the total number of non-flaring ARs that are incorrectly classified as being a member of the positive class (21), TN (True Negative)  denotes the total number of non-flaring ARs that are correctly classified as being a member of the negative class (158), and FN (False Negative) denotes the total number of flaring ARs that are incorrectly classified as being a member of the negative class (10).

\begin{table}[h]
\centering
\caption{The results of flare prediction using the
Zernike moments and the SVM classifier.}
\label{tab2}
\begin{tabular}{llll}
\hline
True Positive (TP) & False Negative (FN) & True Negative (TN) & False Positive (FP) \\ \hline
\hspace*{1.1 cm} 375 &  \hspace*{1.1 cm} 10    &  \hspace*{1.1 cm} 158    & \hspace*{1.1 cm} 21    \\ \hline
\end{tabular}
\end{table}

It is common to assign scores to assess the accuracy of
prediction \citep{Wheatland2005,leka2008}. Several skill scores have been proposed and applied for solar flare predictions. Table \ref{tab3}
represents different skill scores and their related formulae. These metrics are gathered from different papers
on the subject of flare forecasting \citep{woodcock1976,leka2008,Mason2010,Bloomfield2012,bobra2015,Bloomfield2016}.
\afterpage{
\begin{table}[h]
\centering \caption{Definitions of different skill score.}
\label{tab3}
\vspace{0.5cm}
\begin{tabular}{cc}
\hline
Score & Formula \\
\hline
\hline
&\\
Recall (positive and negative) & $\rm{Recall}^{+} = \dfrac{{\rm{TP}}}{\rm{TP} + \rm{FN}}$\\ &\\
&\\& ${\rm{Recall}}^{-} = \dfrac{\rm{TN}}{\rm{TN} + \rm{FP}}$\\ &\\
\hline
&\\
Precision (positive and negative) & ${\rm{Precision}^{+}} = \dfrac{{\rm{TP}}}{\rm{TP} + \rm{FP}}$\\ &\\
&\\& ${\rm{Precision}^{-}} = \dfrac{\rm{TN}}{\rm{TN} + \rm{FN}}$\\ &\\
\hline
&\\
$f_{1}$ score (positive and negative) & $f_{1}^{+} = \dfrac{2 \times \rm{precision}^{+} \times \rm{recall}^{+} }{\rm{precision}^{+} + \rm{recall}^{+}}$\\ &\\
&\\& $f_{1}^{-} = \dfrac{2 \times \rm{precision}^{-} \times \rm{recall}^{-} }{\rm{precision}^{-} + \rm{recall}^{-}}$\\ &\\
\hline
&\\
Accuracy & $\rm{Accuracy} = \dfrac{\rm{TP} + \rm{TN}}{\rm{TP} + \rm{FN} + \rm{TN} + \rm{FP}}$\\
&\\
& \\
\hline
& \\
Heidke Skill Score (HSS) &$\rm{HSS}_{1} = \dfrac{\rm{TP} - \rm{FP}}{\rm{TP} + \rm{FN}}$\\
&\\
\hline
&\\
Heidke Skill Score (HSS)&$\rm{HSS}_{2} = \dfrac{2 \times [(\rm{TP}  \times \rm{TN}) - (\rm{FN}  \times \rm{FP})]}{(\rm{TP} + \rm{FN}) \times (\rm{TN} + \rm{FN}) + (\rm{TP} + \rm{FP}) \times (\rm{TN} + \rm{FP})}$ \\
&\\
\hline
&\\
Gilbert Skill Score (GS)& $\rm{GS} = \dfrac{\rm{TP} - \rm{CH}}{\rm{TP} + \rm{FP} + \rm{FN} - \rm{CH}},$ \\
&\\
&$\rm{CH} = \dfrac{(\rm{TP} + \rm{FP}) \times (\rm{TP} + \rm{FN})}{\rm{TP} + \rm{FN} + \rm{TN} + \rm{FP}}$\\ &\\ \hline
&\\True Skill Statistic (TSS)&$\rm{TSS} = \dfrac{{\rm{TP}}}{\rm{TP} + \rm{FN}} - \dfrac{{\rm{FP}}}{\rm{FP} + \rm{TN}}$\\
&\\
&\\\hline
\end{tabular}
\end{table}
\clearpage
}

\begin{table}[]
\centering
\caption{Performance metrics of the SVM classifier in predicting large flares compared with other authors. We use the first 528 Zernike moments of the magnetogram images. The results of other forecasting methods --
Bobra \#{1} and Bobra \#{2} \citep{bobra2015}; Mason \citep{Mason2010}; Ahmed \citep{ahmed2013}; Barnes \citep{leka2008}; Bloomfield \citep{Bloomfield2012}; Yu \citep{yu2009}; Song \citep{song2009} -- that are described in this table are duplicated from \cite{bobra2015}. Bobra \#{1} and Bobra \#{2} represent the results that are given in the first column of Table 2 and Table 3 of \cite{bobra2015}, respectively.}
\label{tab4}
\begin{tabular}{lccccccccc}
\hline
Metric & \multicolumn{1}{l}{This paper} & \multicolumn{1}{l}{~~~~Bobra \#{1}} & \multicolumn{1}{l}{~~~~Bobra \#{2}} & \multicolumn{1}{l}{Mason} & \multicolumn{1}{l}{Ahmed} & \multicolumn{1}{l}{Barnes} & \multicolumn{1}{l}{Bloomfield} & \multicolumn{1}{l}{~~Yu} & \multicolumn{1}{l}{Song} \\ \hline \hline
\multicolumn{1}{c}{} & 48h                          & 48h                          & 48h                        & 6h                        & 48h                       & 24h                        & 24h                            & 48h                    & 24h                      \\
Recall$^{+}$                            & 0.974                        & 0.714 $\pm$ 0.048                        & 0.869 $\pm$ 0.036           & 0.617                     & 0.677                     & NA                         & 0.704                          & 0.817                  & 0.647                    \\
Recall$^{-}$                            & 0.882                        & 0.989 $\pm$ 0.003                        & 0.947 $\pm$ 0.007           & 0.695                     & 0.994                     & NA                         & NA                             & NA                     & 0.974                    \\
Precision$^{+}$                         & 0.946                        & 0.797 $\pm$ 0.050                        & 0.501 $\pm$ 0.041           & 0.008                     & 0.877                     & NA                         & 0.146                          & 0.831                  & 0.917                    \\
Precision$^{-}$                         & 0.940                        & 0.983 $\pm$ 0.003                        & 0.992 $\pm$ 0.002           & 0.998                     & 0.980                     & NA                         & NA                             & NA                     & 0.860                    \\
$f_{1}^{+}$                                & 0.959                        & 0.751 $\pm$ 0.032                        & 0.634 $\pm$ 0.033           & 0.015                     & 0.764                     & NA                         & 0.242                          & NA                     & 0.758                    \\
$f_{1}^{-}$                                & 0.910                        & 0.986 $\pm$ 0.002                        & 0.969 $\pm$ 0.003           & 0.819                     & 0.987                     & NA                         & NA                             & NA                     & 0.913                    \\
Accuracy                                     & 0.945                        & 0.973 $\pm$ 0.003                        & 0.943 $\pm$ 0.006           & 0.694                     & 0.975                     & 0.922                      & 0.830                          & 0.825                  & 0.873                    \\
$\rm{HSS}_{1}$                                   & 0.919                    & 0.528 $\pm$ 0.062                 & -0.008 $\pm$ 0.142           & -78.9                     & 0.581                     & 0.153                      & NA                             & NA                     & 0.588                    \\
$\rm{HSS}_{2}$                                   & 0.871                   & 0.737 $\pm$ 0.034                  & 0.606 $\pm$ 0.035           & 0.008                     & 0.751                     & NA                         & 0.190                          & 0.650                  & 0.676                    \\
GS                                           & 0.771                        & 0.585 $\pm$ 0.043                        & 0.436 $\pm$ 0.036           & 0.004                     & 0.601                     & NA                         & NA                             & NA                     & 0.510                    \\
TSS                                          & 0.856                        & 0.703 $\pm$ 0.047                        & 0.817 $\pm$ 0.034           & 0.312                     & 0.671                     & NA                         & 0.539                          & 0.650                  & 0.620                    \\ \hline
\end{tabular}
\end{table}

Table \ref{tab4} lists the prediction metrics achieved by the present algorithm compared to the scores of other works.
Bobra \& Couvidat (2015) provided two tables (Tables 2 and 3 therein) to compare the values of skill scores obtained by
different forecasting methods. Other than the second column of Table \ref{tab4}, all other columns are
a copy of Tables 2 and 3 of \cite{bobra2015}.

\cite{leka2008} concluded that even if different databases are used for prediction, comparison of the skill scores for different methods is meaningful. However, unless the datasets
are identical, there is no completely meaningful comparison between two or more different methods that one could make.
Hence, one should not
consider the results of Table \ref{tab4} as an absolute reference for comparison between the methods.

The second column of Table \ref{tab4} lists the skill scores of the present algorithm for the classification process (see the two supplementary electronic tables). The third and fourth column of Table \ref{tab4} list the performance metrics
achieved by \cite{bobra2015},
for specifically tuned SVMs which result in the highest HSS$_{2}$, and TSS scores respectively.
Their method
was demonstrated to predict large solar flares within 48 hours before occurrence
with a TSS score of $0.817$. The second-highest TSS in the table belongs to \cite{ahmed2013} which is $0.671$.  As Table \ref{tab4} shows, the TSS
score achieved in the present work is 0.856.
Also, the highest HSS$_{2}$ score amongst all other previous methods, given by \cite{ahmed2013}, is $0.751$
and the second-highest HSS$_{2}$ score belongs to \cite{bobra2015}, which is $0.737$. The HSS$_{2}$ score
attained with the present method is 0.871.
Another metric of interest here is the Recall$^{+}$. As it is shown in Table \ref{tab3}, the
Recall$^{+}$ score is associated with the number of FNs and TPs, which characterize the ability of the classifier
to achieve the least number of FNs. The reason for this interest is that if a positive event is falsely reported as
being a negative one, the resulting costs for this lack of accuracy in prediction could be devastating.
Mis-prediction of negative events (i.e. false positive, FP) may only require, for example, powering off
a power plant, or rotating a satellite's shields towards the Sun, but when it is reported
to an astronaut in deep space that they are unlikely to be hit by a large flare within some time, the consequence of error is more serious.
The highest Recall$^{+}$ score among former methods is $0.869$ due to \cite{bobra2015}, and the second-highest Recall$^{+}$ is $0.817$, due to \cite{yu2009}. The Recall$^{+}$ score gained by the present method
is 0.974.

\section{Discussion} \label{sec:discussion}

Here, we propose a method based on the properties of ZMs (Zernike
Moments) of magnetogram images for ARs (Active Regions), and the
Support Vector Machine for prediction of large (M and X type)
solar flares.

Previous methods have used a few parameters extracted from AR 
magnetograms (e.g. the total unsigned current helicity, the total 
magnitude of the Lorentz force, the total photospheric magnetic free 
energy density, and the total unsigned vertical current) as the basis 
for classification \citep[e.g.][]{leka2003,bobra2015,all-clear}. One may ask, is it possible that two
different magnetic fields yield the same  values for above
mentioned parameters? Suppose we have two arbitrary three dimensional vector
magnetic field magnetograms observed at the solar photosphere ($z = 0$), ${\bf B}_1(x,y)$ and
\begin{equation} \label{eq8}
{\bf B}_2(x,y)={\bf B}_1(x,y)+\nabla\phi(x,y).
\end{equation}
These two magnetic fields lead to the
same total current helicity
\begin{eqnarray} \label{eq9}
\sum B_{2z}J_{2z} & = & \sum \frac{1}{\mu_{0}}(\nabla \times {\bf B}_{2})_{z} B_{2z} \nonumber \\ & = & \sum \frac{1}{\mu_{0}} (\nabla \times {\bf B}_{1})_{z} B_{1z} \nonumber \\  &=& \sum B_{1z}J_{1z},
\end{eqnarray}
 and the same total flux
 \begin{equation} \label{eq10}
\sum B_{1z}(x,y) dxdy=\sum B_{2z}(x,y) dxdy.
\end{equation}
Since the total free energy density and the total Lorentz force are both proportional to $B^{2}$, applying an additional
constraint,
\begin{equation}
2\frac{\partial\phi}{\partial x}B_{1x}+\left(\frac{\partial\phi}{\partial x}\right)^2 = -2\frac{\partial\phi}{\partial y}B_{1y}-\left(\frac{\partial\phi}{\partial y}\right)^2,
\end{equation}
results in
\begin{equation}
B_{1}^{2} = B_{2}^{2},
\end{equation}
and hence the total free energy density and the total Lorentz force for the two magnetic fields are the same. Assume that
 these two vector magnetic  field represent the photospheric magnetic field for two arbitrary ARs.
  It may happen that one of the magnetic fields corresponds to a flare productive AR and the other corresponds to a non-flaring AR. In this case a classification process based on helicity, total flux, total free energy, and total Lorentz force will not discriminate between the two ARs, since these two different magnetic fields have the same values for above mentioned parameters. In other words, there could be two different vector magnetic fields for two ARs having an identical vector in the feature space. This can obviously affect the results of the classification. It can be seen that the ZMs for these two magnetic fields
  given by Equation (\ref{eq3}) represent two different  sets of values\footnote{This example is not intended to
  be realistic: two real vector magnetograms will not have identical values of $J_{z}$ and $B^{2}$. However, the example demonstrates the principle that two different magnetic fields may have the same values of these parameters.}.

Moreover, as discussed in \cite{all-clear}, performance comparisons between different flare forecasting methods based on extracting a few parameters out of AR magnetograms indicate that there is no clearly superior method, and it was pointed out that this might be due to correlations between the parameters. Also, the methods were found to have a rather weak performance in achieving high positive skill scores.
An advantage of the present method is that the Zernike moments provide unique information as a basis for classification of an AR by comparison with a few global parameters (e.g., total flux, current helicity etc).
Further, the present method is demonstrated to be able to predict solar flares with a small number of FNs rather than just reducing the number of FPs. This has important practical consequences for reducing the costs of errors in prediction (e.g. Bobra \& Couvidat 2015).
\\

We thank NASA/SDO, the HMI Science Team, and NOAA for the availability of the SDO/HMI data and the GOES flare catalogue used here.

\allauthors

\listofchanges

\section*{\textbf{Appendix}}
\subsection{The Support Vector Machine}
The purpose of the SVM 
 classifier is to find a decision boundary with
 a margin as large as possible, to reduce the
classification error. Suppose that $D$ is a binary-class training
set with $N$ data points in the $l$-dimensional feature $(x_{1},
..., x_{l})$ space, that is
\begin{equation}
D = \bigg \{(x_{i}, y_{i}) | x_{i} \in R^{l}, y_{i} \in \{-1 ,
+1\})\bigg \},~~i=1,{\cdots}, N.
\end{equation}

Constructing a decision boundary, which is a separating
hyperplane in a high-dimensional space, SVM can segregate classes.
This hyperplane is given by
\begin{equation}
\textbf{w}\cdot\boldsymbol{\Phi}(\textbf{x}) + b = 0,
\end{equation}
where
\begin{equation}
\boldsymbol{\Phi} : R^{l} \rightarrow R^{L}, ~~~~~~~~ L \geq l,
\end{equation}
and where $\textbf{w}$ and $b$ are a weight vector and bias,
respectively. $\boldsymbol{\Phi}(\textbf{x})$ is a linear or nonlinear vector
function that maps each data point $\textbf{x}_{i}$ into the feature space in
high-dimensional space. These parameters, namely $\textbf{w}$ and $b$, can be computed by
solving the following optimization problem:
\begin{eqnarray}
& & \min_{\textbf{w},b,\xi} ~~ \dfrac{1}{2} ||\textbf{w}^{2}|| +
c \sum_{i=1}^{N}{\xi_{i}},
\end{eqnarray}
subject to (the constraint):
\begin{eqnarray}
y_{i}(\textbf{w}\cdot\textbf{x}_{i} +b) \geq 1 - \xi_{i}, ~~~~~~~~ \xi_{i}
\geq 0,
\end{eqnarray}
where $\xi_{i}$ and $c$ are the error value for the decision boundary and
the regularization parameter, respectively. The regularization
parameter controls the trade-off between the margin width and model
complexity and is determined by the user. The equations given above
can be converted into the following dual form:
\begin{eqnarray}
& & \max_{\alpha} ~~ \sum_{i=1}^{N}\alpha_{i} - \dfrac{1}{2}
\sum_{i=1}^{N}\sum_{j=1}^{N}
y_{i}y_{j}\alpha_{i}\alpha_{j}K(x_{i},x_{j}),
\end{eqnarray}
subject to
\begin{eqnarray}
  \sum_{i=1}^{N}y_{i}\alpha_{i} = 0, ~~~~~~~~ \alpha_{i} \geq 0, ~~~~~~~~ \forall i : 0 \leq i \leq c,
\end{eqnarray}
where $\alpha_{i}$ is the Lagrange multiplier corresponding to the
$i^{\rm th}$ training sample and $K(x_{i}, x_{j})$ is a Kernel
function which maps the input vectors into a suitable feature space to achieve a better representation.
So, we have $K(x_{i}, x_{j}) =
\boldsymbol{\Phi}(\textbf{x}_{i}) \cdot \boldsymbol{\Phi}(\textbf{x}_{j})$. This is a
constrained optimization problem and it can be solved by a
Lagranian multiplier method. The output of SVM for each input
data point is equal to
\begin{eqnarray}
& & y(x) = \rm{sgn}[\textit{f}(\textit{x})],
\end{eqnarray}
where
\begin{eqnarray}
f(x) = \sum_{i=1}^{N}y_{i}\alpha_{i}K(x_{i}, x_{j}) + b.
\end{eqnarray}
Usually after training the SVM, the value
of Lagrange multiplier is zero for many training points. Support vectors are input vectors
that just touch the boundary of the margin \citep[see e.g.][]{qu2003,theodoridis2009,hsu2011}.
\end{document}